\documentclass[a4paper,11pt]{article}
\pdfoutput=1 
\usepackage{jcappub} 
\usepackage{fontenc} 
\usepackage{tabularx}
\usepackage{hhline}

\def\ts{\rm TS}

\title{Discovery of a New Galactic Center Excess Consistent with Upscattered Starlight}

\author[a,]{Kevork N.\ Abazajian,}
\author[a]{Nicolas Canac,}
\author[a,b]{Shunsaku Horiuchi,}
\author[a]{Manoj Kaplinghat,}
\author[a]{and Anna Kwa}

\affiliation[a]{Center for Cosmology, Department of Physics and Astronomy, University of California, Irvine, Irvine, California 92697 USA}
\affiliation[b]{Center for Neutrino Physics, Department of Physics, Virginia Tech, Blacksburg, Virginia 24061, USA}

\emailAdd{kevork@uci.edu}
\emailAdd{ncanac@uci.edu}
\emailAdd{horiuchi@vt.edu}
\emailAdd{mkapling@uci.edu}
\emailAdd{akwa@uci.edu}

\abstract{We present a new extended gamma ray excess detected with
the Fermi Satellite Large Area Telescope toward the Galactic Center
that traces the morphology of infrared starlight emission. Combined with
its measured spectrum, this new extended source is approximately consistent with
inverse Compton emission from a high-energy electron-positron
population with energies up to about 10 GeV. Previously detected
emissions tracing the 20 cm radio, interpreted as bremsstrahlung
radiation, and the Galactic Center Extended emission tracing a
spherical distribution and peaking at 2 GeV, are also detected. We
show that the inverse Compton and bremsstrahlung emissions are likely
due to the same source of electrons and positrons. All three extended
emissions may be explained within the framework of a model where the
dark matter annihilates to leptons or a model with unresolved
millisecond pulsars in the Galactic Center.}
          
\begin{document}
\maketitle
\flushbottom

\section{Introduction}
\label{sec:intro}
The Fermi Gamma Ray Space Telescope
Large Area Telescope (Fermi LAT) has observed with unprecedented
detail the ``heart of darkness'' of our Galaxy: its gravitational
center. The past few years have revealed that there are a large number
of new point sources \cite{Nolan2012} as well as new diffuse
emission~\cite{YusefZadeh:2012nh}. In addition, a large extended
source, the Galactic Center Extended (GCE), has been detected by a
number of groups
\cite{Goodenough:2009gk,Hooper:2010mq,Hooper:2011ti,Abazajian:2012pn,Daylan:2014rsa}
and is robust to uncertainties in the diffuse emission foregrounds in
the region
\cite{Gordon:2013vta,Abazajian:2014fta,Zhou:2014lva,Calore:2014xka}.
The high-energy radiative processes that produce gamma rays are often
commensurate with production or acceleration of related relativistic
charged particle cosmic rays.  Astrophysical processes include
diffusive shock acceleration, magnetic reconnection, ``one-shot''
acceleration across high-voltage electric fields, and many other
possibilities.  Another source that can produce both cosmic rays and
gamma rays are the products from candidate dark matter particle
annihilation or decay. The significance of the Galactic center as a
bright source for dark matter annihilation photons and cosmic rays has
been known for some time \cite{Bergstrom:1997fj}.

High-energy charged particles, deposited either from astrophysical sources or dark matter 
annihilation, experience various propagation and energy-loss 
processes in the Galactic Center region.
There has been recent work discussing how bremsstrahlung and inverse
Compton (IC) effects could alter the prompt spectra coming from dark
matter annihilation \cite{Cirelli:2013mqa,Lacroix:2014eea}. What we
show here, for the first time, is that these separate components---prompt, 
bremsstrahlung and IC---can be separated with morphological as
well as spectral information. In particular, we report the discovery of a new extended
component of the gamma-ray emission toward the Milky Way Galactic Center
that is spectrally and morphologically consistent with a population of 
electron-positron ($e^\pm$) cosmic rays producing gamma rays by
upscattering starlight through the IC process. Secondly, 
we confirm the  presence of an emission consistent with bremsstrahlung 
radiation, and find that this emission can be produced by the same population 
of $e^\pm$ impinging on the high-density gas in the Galactic Center. Lastly, 
we confirm the presence of a GCE source (peaking around 2 GeV) that has 
a centrally-peaked morphology consistent with dark matter annihilation. We show that 
the IC, bremsstrahlung and GCE components could all have originated from 
the products of dark matter annihilation. This explanation is not unique in that an 
unresolved population of millisecond pulsars or two independent astrophysical 
sources could produce these signals.

\section{Methods and Results}
\label{sec:methods}
We use Fermi Tools version \verb|v9r33| to
study Fermi LAT data from August 2008 to June 2014 (approximately 70
months of data). We use Pass 7 rather than Pass 7 Reprocessed
instrument response functions since the diffuse map associated with the
latter have strong caveats for use with new extended sources. Our procedure is similar to those
described in Ref.~\cite{Abazajian:2014fta}. We simultaneously fit the
amplitude and spectrum of point sources from the 2FGL
catalog~\cite{Nolan2012}, plus four other point sources in the ROI,
as described below, in our region of interest (ROI) $7^\circ\times
7^\circ$ around the GC centered at $b=0, \ell=0$.  We use $0.2 -
100\ {\rm GeV}$ photons in 30 logarithmically-spaced energy bins. To
enhance spatial resolution, we use \verb|ULTRACLEAN|-class photons
binned in an Aitoff projection into pixels of $0.1^\circ\times
0.1^\circ$.

We include the 20 cm radio template as a tracer of gas to account for
the bremsstrahlung emission as has been done
previously~\cite{YusefZadeh:2012nh,Macias:2013vya,Abazajian:2014fta}.
To test the possibility of IC emission from starlight due to this same
population of $e^\pm$, we use the 3.4 $\mu\rm m$ template for stellar
light from the WISE mission~\cite{Wright2010}.  Among the templates
tested, this had the least obscuration of stellar light in the ROI; the 
results from other templates studied are discussed later. 
Our goal in using the 3.4 $\mu\rm m$ template is to test whether the 
IC component's morphology might be approximated by it; 
we do not presume that this template is an exact morphological 
description of the putative IC emission. 
As an example, if the diffusion length of $e^\pm$ is significantly less than
the ROI dimensions, then the IC emission will track the morphology of the 
$e^\pm$ source more.

We use a $14^\circ\times 14^\circ$ template because of the broad PSF
of Fermi-LAT producing contributions outside of the ROI, particularly
at low energies consistent with the IC photons. As in
Ref.~\cite{Abazajian:2014fta}, we also include the New Diffuse (ND)
map whose intensity increases with angle away from the GC, which was
interpreted as accounting for additional gas not captured in the 20 cm
map.  We have optimized the morphology of the GCE excess and ND
templates to their best-fit profiles, as in
Ref.~\cite{Abazajian:2014fta}. To optimize the GCE excess, we used
templates of $\rho(r)^2$ projected along the line-of-sight with
$\rho(r) \propto r^{-\gamma}(r+r_s)^{-(3-\gamma)}$ and found that
$\gamma=1$ provided the best-fit. The best-fit new diffuse template
increases with projected distance from the Galactic Center, $\theta$, as
$\theta^{0.3}$. It is worth noting that our GCE template is somewhat
less steep than found previously, $1.1-1.4$ for the inner slope of the
density profile~\cite{Abazajian:2014fta,Gordon:2013vta,Daylan:2014rsa}.

In our analysis, we include two previously dicovered point sources,
1FGL J1744.0-2931c and ``bkgA'' \cite{YusefZadeh:2012nh}, and 
furthermore discover two new point sources, PS1 and PS2, 
at $\ell,b$ of $356.616^\circ,1.009^\circ$ and
$356.829^\circ,-0.174^\circ$ with large test statistic (TS)
values of 168 and 140, respectively\footnote{$\ts \equiv 2\Delta\ln{\mathcal{L}}$, where $\Delta\mathcal L$ is the difference of the best fit likelihood with and without the source. For point sources, a value of $\rm TS = 25$ is detected at a significance of just over $4\sigma$ \cite{Nolan2012}.}. PS1 has a spectrum consistent
with a Log-Parabola, is near numerous X-ray and radio sources
including, e.g., 1RXS J173331.6-311522, and could be in the plane of
the Galaxy or extragalactic. PS2 has a spectrum consistent with a
power law, and is near the supernova remnant G356.8-00 and compact
radio source G356.9+0.1~\cite{RoyRao2002}. They are at the edge of our
ROI, and our conclusions regarding the IC, bremsstrahlung and GCE sources 
are not qualitatively affected by their inclusion. We choose to keep them 
in our models.

\begin{figure}[ht!]
\centering
\includegraphics[width=1.95truein]{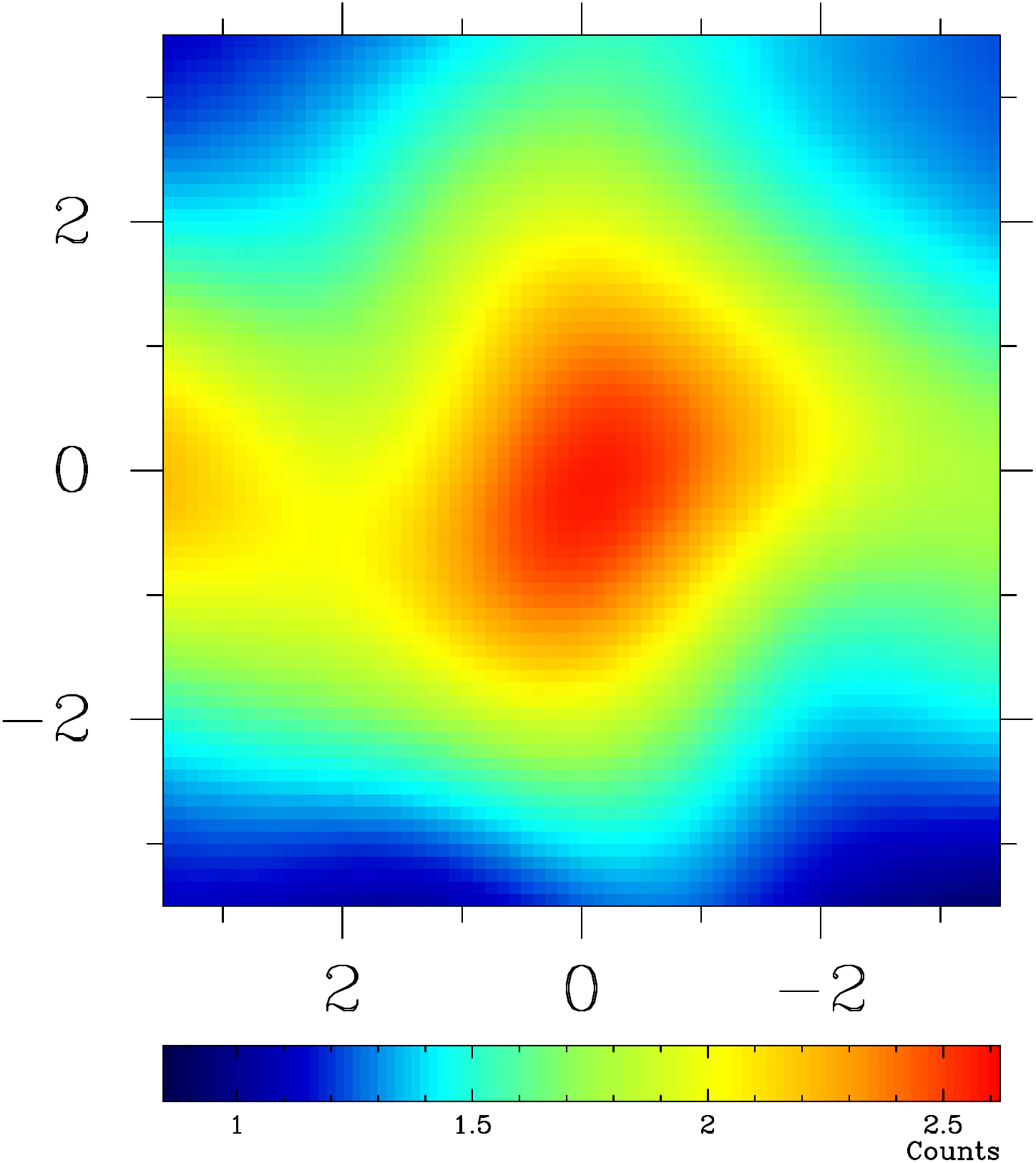}
\includegraphics[width=1.95truein]{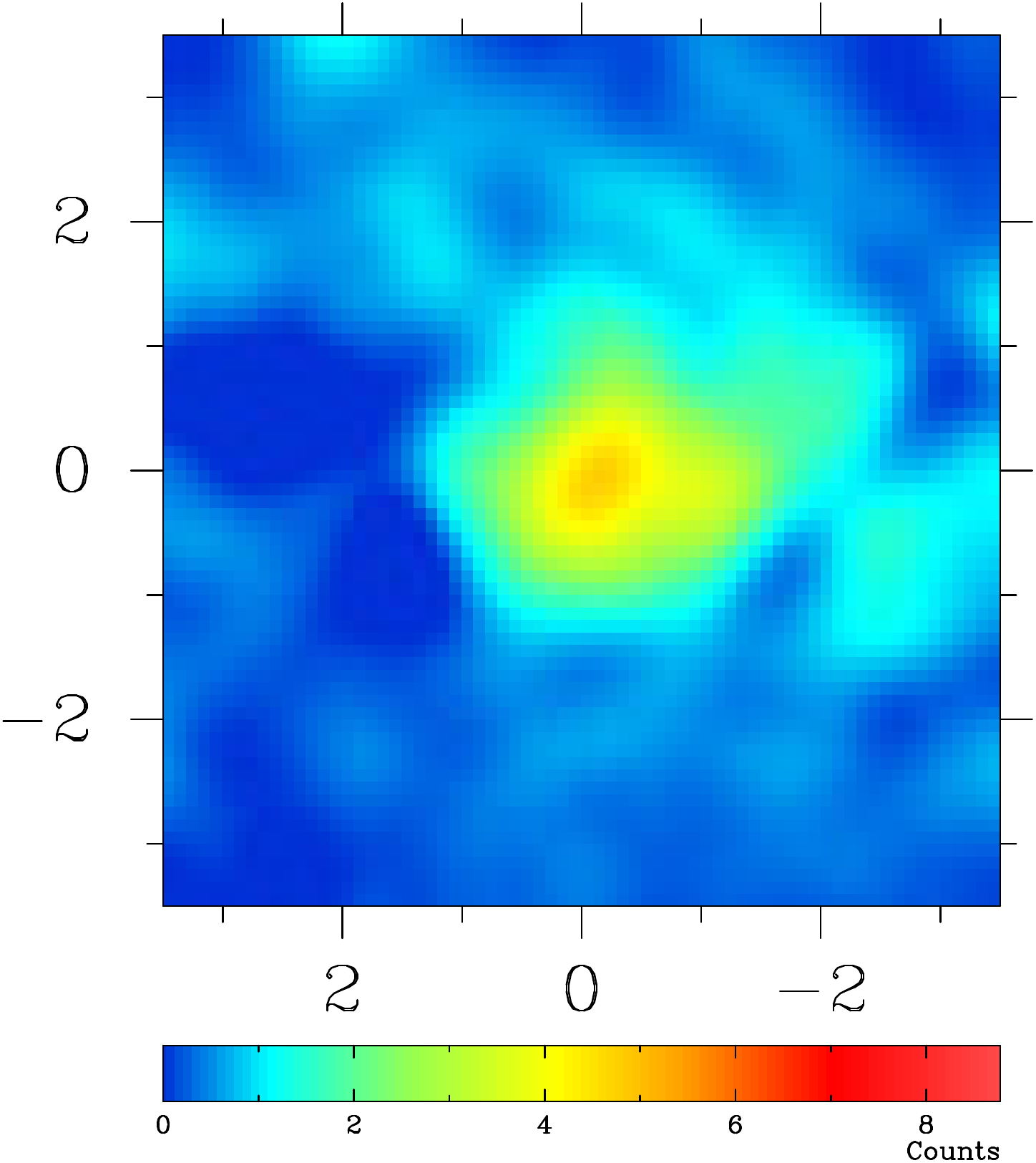}
\includegraphics[width=1.95truein]{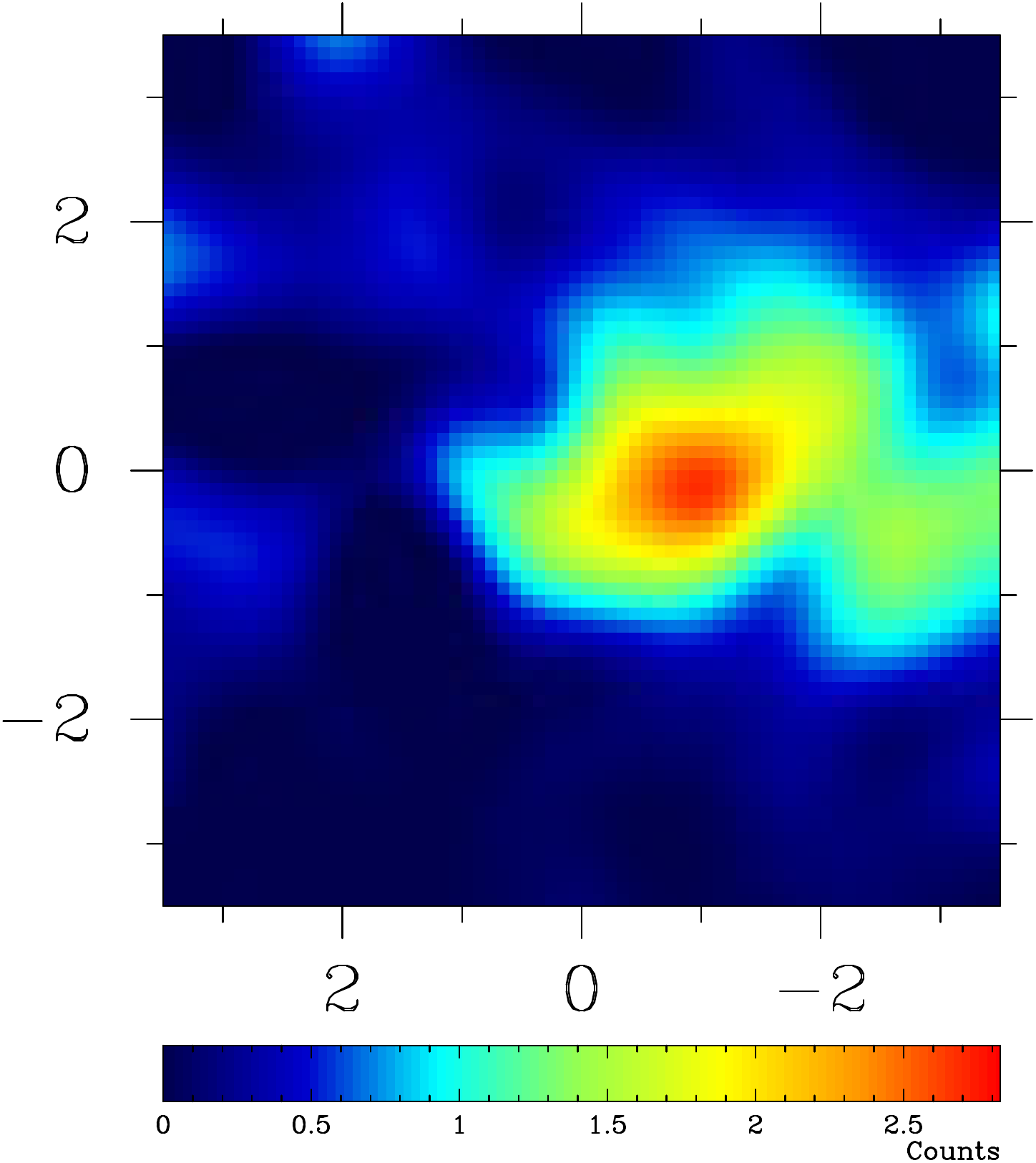}\\

\includegraphics[width=1.95truein]{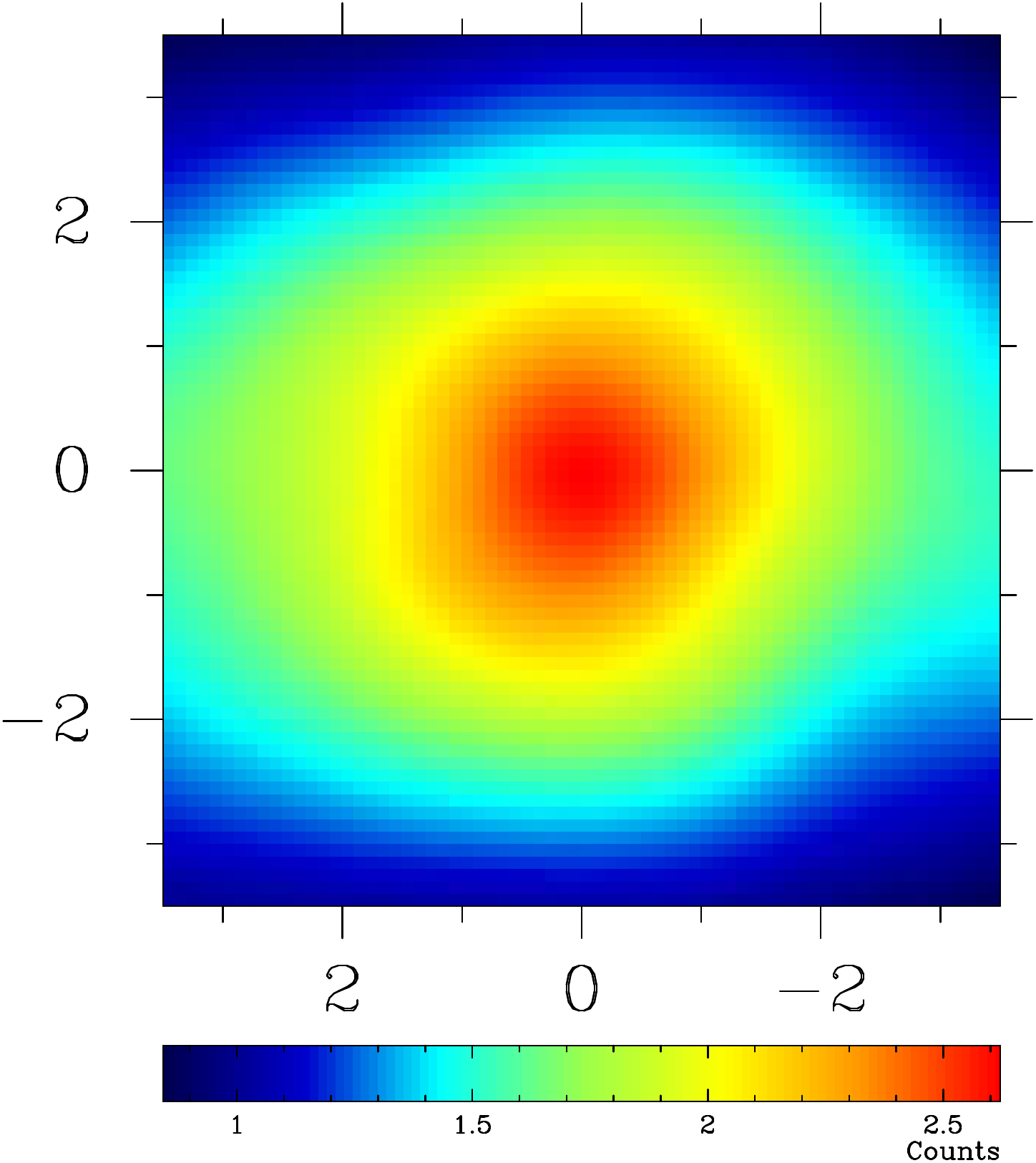}
\includegraphics[width=1.95truein]{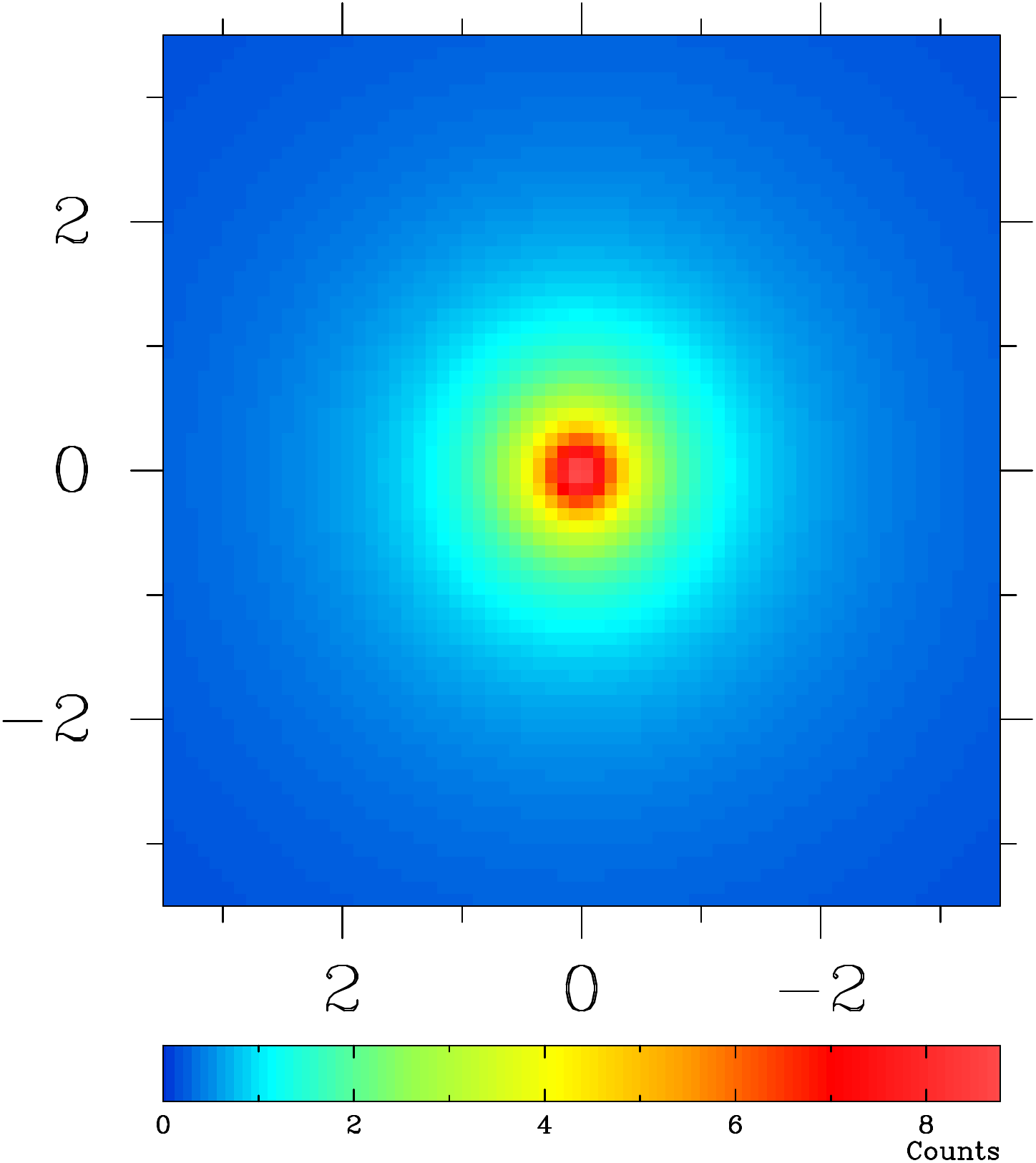}
\includegraphics[width=1.95truein]{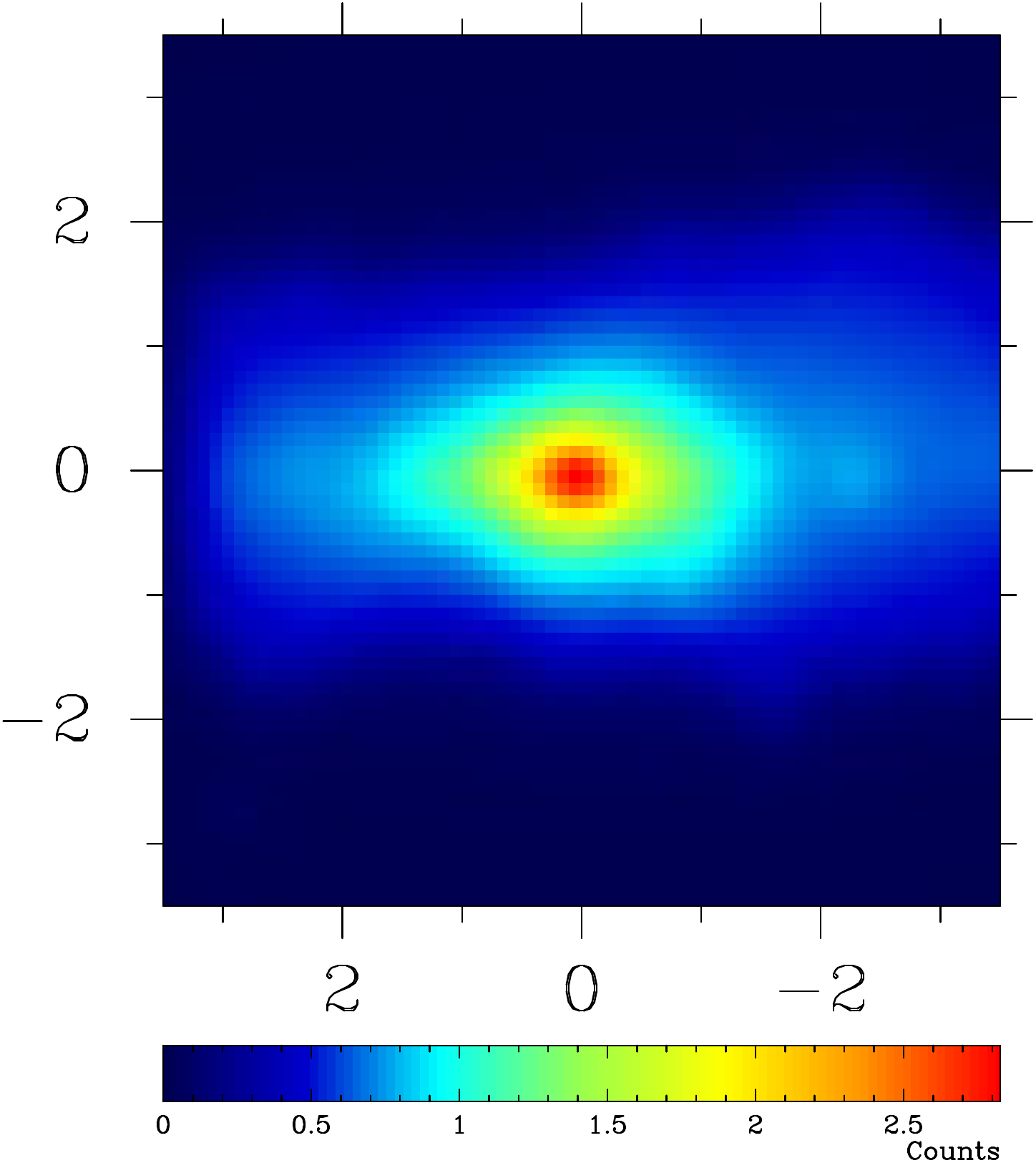}
\caption{Shown in the left column are the residual
  photons (top) and best fit model (bottom) associated with the
  projected interstellar radiation field template in its peak intensity
  bin, $0.303{\rm\ GeV} < E_\gamma < 0.372\rm\ GeV$, where 
  the residual map has been smoothed with a Gaussian of 
  $\sigma = 0.9^\circ$ (to roughly account for the point-spread 
  function). The middle column shows the residual photons (top) and best
  fit model (bottom) associated with the projected dark matter density 
  squared template in its peak intensity bin, $1.59{\rm\ GeV} <
  E_\gamma < 1.95\rm\ GeV$, where the residual map has been 
  smoothed with a Gaussian of $\sigma = 0.4^\circ$. The right column shows the residual photons (top) and best fit model (bottom) 
  associated with the 20 cm radio map in the same energy bin and 
  with the same smoothing as the middle row. Residual and model 
  maps have the same color scale for each row. This analysis used 
  {\tt ULTRACLEAN}-class photons.\label{spatialmap}}
\end{figure}

\begin{figure}[t]
\centering
\includegraphics[width=6truein]{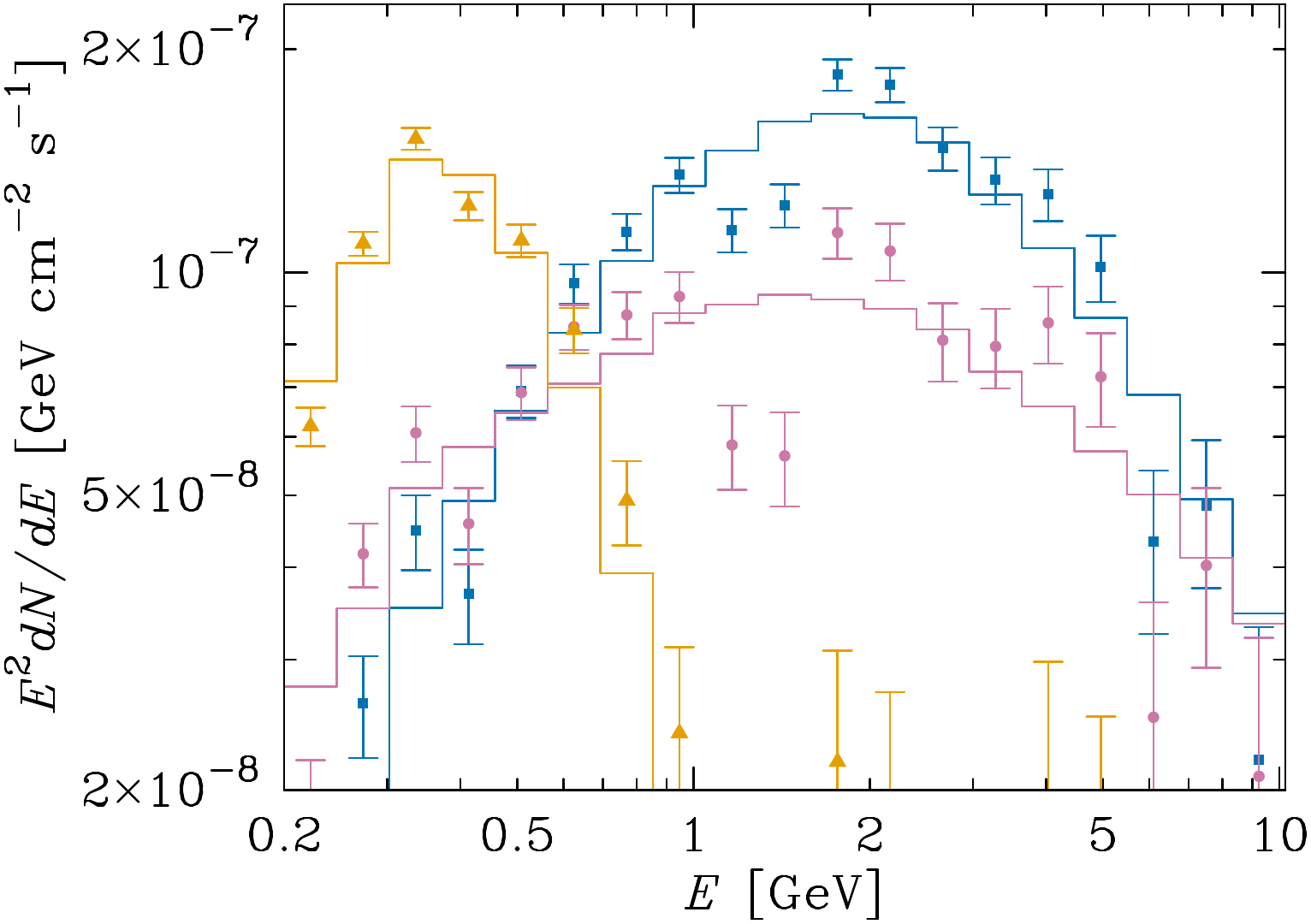}
\caption{The residual spectra (points with errors) and best-fit model spectra 
for the projected interstellar radiation field (golden triangle), gas (pink circle), and 
dark matter density squared (blue square) templates.  {\tt ULTRACLEAN} class photons 
are used for this analysis.\label{spectra}}
\end{figure}

All the 4 extended sources (GCE, ND, IC, Bremsstrahlung) were given 
generic log-parabola spectral forms with four free parameters each. 
We detect the WISE 3.4 $\mu\rm m$ template at very high
significance of $\ts = 197.0$. The previously studied sources were
also detected at high significance. The GCE was detected with $\ts =
207.5$, bremsstrahlung was detected with $\ts = 97.2$.  These
sources and best fit models are shown in Fig.~\ref{spatialmap},
and the resulting residual spectra and best-fit log-parabola models are 
shown in Fig.~\ref{spectra}.


In addition to the IC and Bremsstrahlung signatures of a population of
high energy $e^\pm$, the 20 cm radio emission is also consistent with
the synchrotron emission from the same population of electrons with
correlated implications for the ionization and temperature of the
molecular gas \cite{YusefZadeh:2012nh}. The fact that the
bremsstrahlung emission traces the 20 cm (synchrotron) map indicates
that the magnetic field is frozen into the gas. When we replaced the
20 cm map with a CO map, which contains dense molecular structures
along the plane, the bremsstrahlung excess was not detected.

We again emphasize that we do not expect the WISE template to be an exact morphological description for the proposed IC emission from GCE-associated electrons. However, the WISE template's high TS value does indicate that it is indeed a reasonable approximation for the IC component. We tested two other templates for the IC component. With a $100~\mu m$ dust template
map \cite{Schlegel:1997yva}, we were able to detect essentially the
same IC spectrum with almost the same TS value. This indicates that
the IC emission traces a disky template (thicker than the
bremsstrahlung emission) but that there are considerable uncertainties
in determining the correct morphology due to the poor angular
resolution at energies below 500 MeV. With a 2MASS J-band ($1.2~\mu
m$) template \cite{Skrutskie:2006wh}, the significance of the IC
detection was much lower (TS = 98.4); this is likely due to the large
variable dust attentuation evident in the J-band map.  For both , we observe the notable feature of the IC spectral cutoff at $\sim$1 GeV. This is a distinctive spectral feature of the IC excess that distinguishes it from the GALPROP-calculated IC emission contained within the diffuse background model.
 
To test the robustness of our results, we repeat our analysis using different diffuse backgrounds generated using the GALPROP code \cite{galprop}. We tested two models in the extreme case where parameters were chosen with the intent to increase the IC emission predicted in the diffuse background, thus increasing the possibilty that some or all of the excess might be absorbed into the background. For the first of these extreme cases, we chose a model with a very low diffusion parameter, which results in increased IC emission at lower energies. For the second extreme case we tested a model with a factor of 1.5 increase in optical and IR ISRF normalizations. We also tested two models with more standard parameters that were found by \cite{Calore:2014xka} to be good fits to the data in the inner galaxy. In all tests, the IC and GCE components were recovered with very similar spectra and uncertainties as when the Pass 7 model background was employed. The bremsstrahlung component was similar in most test cases; the only notable difference was that when employing a background model with extremely low diffusion coefficient, the bremsstrahlung spectrum did not show a cut-off as in Fig.~\ref{spectra}, but it still had the same flux at GeV energies. These findings support our claim that the excesses are robust, and not an artifact of using the Pass 7 diffuse model. Interested readers may find the details of these diffuse background model checks in the following appendix.

We also test the dependence of our main results on the extended source templates included when modelling the data. We confirm that the spectra of the IC and GCE components remain more or less unchanged when the data is fit without the MG and ND components. We also substitute the HESS collaboration's $\sim$TeV residual map of the Galactic Ridge \cite{HESScollab:2006} in place of the 20 cm map used as the bremsstrahlung template, and find that the GCE and IC spectra again remain very consistent with the results shown in Fig. \ref{spectra}, while the HESS map spectrum has a slightly lower energy cutoff (around 1 GeV) than the 20 cm map spectrum (around 2--4 GeV) but with similar peak normalization. Details of these checks may also be found in the appendix.

\section{Interpretation and Discussion}
We discuss here how the
detected IC emission is consistent with arising from the same
population of $e^\pm$ as that producing the bremsstrahlung
emission. In addition, we show that the GCE, IC, and bremsstrahlung 
emission can all arise from dark matter annihilation to leptons.

Apart from the dark matter interpretation \cite{Hooper:2011ti}, the GCE 
has been proposed to be emission from millisecond
pulsars (MSPs) \cite{Abazajian:2010zy,Abazajian:2012pn}, episodic hadronic
\cite{Carlson:2014cwa}, or episodic leptonic cosmic ray injection
\cite{Petrovic:2014uda}. Pulsars have the right conditions to produce energetic 
$e^\pm$ cosmic rays~\cite{GoldreichJulian1969} and hence, in
principle, MSPs could explain all three excesses: the GCE 
excess due to the gamma-ray emission from their outer
magnetosphere, and the IC and bremsstrahlung resulting from the
$e^\pm$ that are produced along with gamma-rays in 
cascades~\cite{1987ICRC....2...92H}. Hadronic emission is less
promising because it has trouble with the observed symmetry of the 
GCE . The IC emission must arise from a leptonic channel, perhaps 
secondary $e^\pm$ produced due to hadronic interactions or a separate channel 
such as leptonic cosmic ray emission from star formation activity. 

The fact that bremsstrahlung and the GCE spectrum could originate 
from a broken spectrum of $e^\pm$ resulting from dark matter
annihilation has been discussed previously~\cite{Linden:2011au,YusefZadeh:2012nh}. 
Below we argue that the bremsstrahlung and IC spectra may naturally be related 
to the same $e^\pm$ population, which in turn could be connected to the origin
of the GCE excess.

Let us consider a population of $e^\pm$ with energy $E_c$.  
The resulting IC photons have typical energies of $(E_c/m_e)^2 hc/(1~\mu {\rm m})$, 
where we have used the fact that the spectral energy distribution of the interstellar radiation field peaks around a micron. Assuming $E_c=10~{\rm GeV}$ results in an IC spectrum that cuts off rapidly by 1-2 GeV. The bremsstrahlung spectrum for the same population is broader in energy and extends up to $E_c$. Both these predictions are qualitatively consistent with the spectra shown in Fig.~\ref{spectra}. To test the consistency of the spectra with this simple picture further, we build a simplified model of diffusion and energy loss in the Galactic Center. 

\begin{figure}[t!]
\centering
\includegraphics[width=6truein]{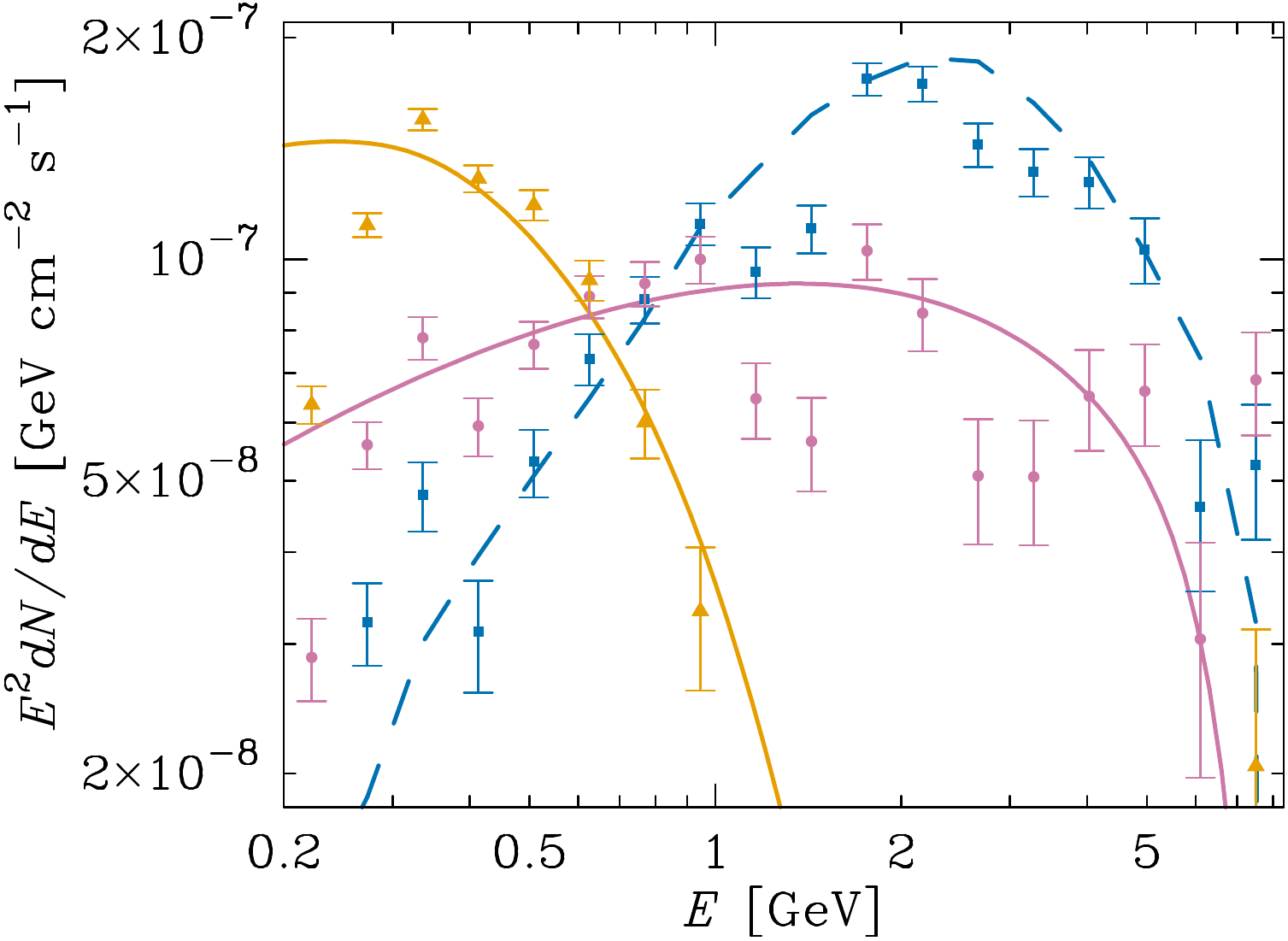}
\caption{Shown here is an example 8 GeV dark matter annihilation
  model with equal branching to all charged leptons,
  $e^\pm,\mu^\pm,\tau^\pm$, with the residual spectra of the prompt
  GCE (blue square), IC (golden triangle), and bremsstrahlung (pink
  circle) sources. The blue (dashed) GCE spectrum is is determined by
  the particle mass and annihilation rate fit to the observations. The
  solid {\em predicted} resultant spectra for this annihilation
  channel's IC (golden) and bremsstrahlung (pink) cases are in solid
  lines.  {\tt ULTRACLEAN} class photons are used for this
  analysis.\label{spectraDM}}
\end{figure}

The $e^\pm$ in the products created by dark matter annihilation lose energy through 
three distinct process~\cite{Colafrancesco:2005ji}: (1) IC, which 
leads to upscattering of the interstellar radiation field (ISRF) photons, (2) bremsstrahlung (Br) 
radiation off the gas, and (3) synchrotron radiation in the Galactic magnetic field. We focus on the
first two components in this {\em letter}. The differential flux of photons for 
these two components may be written as, 
\begin{equation}
E\frac{dN_{\rm IC,Br}}{dE} = \int_{\rm FOV} \frac{d\Omega}{4
  \pi} \int_{\rm LOS} d\ell \int_{E_{\rm min}}^{m} dE_e
    \frac{dn_e}{dE_e} \frac{dP_{\rm IC,Br}}{dE}\,
\end{equation}
where FOV and LOS indicate integration over the field-of-view and line-of-sight 
respectively, $dP_{\rm IC}/dE$ and $dP_{\rm Br}/dE$ are the differential
power emitted per electron due to IC and bremsstrahlung
processes. For bremsstrahlung, we include energy losses from atomic H and He. 
To get the source energy distribution of electrons, positrons and gamma rays, we 
use the software {\sc PPPC4DMID}~\cite{Cirelli:2010xx}. 
The number density of electrons and positrons per unit energy,
$dn_e/dE_e$, is computed after including diffusion and energy losses
according to the prescriptions in Refs.~\cite{Delahaye:2007fr, Cirelli:2013mqa}. 

To propagate the $e^\pm$, we assume a spatially constant diffusion coefficient 
$K(E)=K_0 E^\delta$, with $K_0$ and $\delta$ set to the "MED" model~\cite{Delahaye:2007fr} 
(often used for diffusion in the local neighboorhood). The diffusion process is largely
unconstrained in the Galactic Center and variations away from the assumed parameters
have significant effects on the magnitude and spatial profile of the IC and 
Br signals. Also, the analytic description for diffusion does not 
allow for spatially-varying energy loss terms and we have assumed 
average values for the ISRF energy density and the gas density to create the model spectra for 
comparison. These choices, however, serve to illustrate our two main points that (a) the IC and bremsstrahlung spectrum could be due to the same lepton population, and (b) a single mechanism 
could explain the morphologies, strengths and spectra of the three distinct Galactic Center 
extended excesses. 

For the average gas density and magnetic field strength, we assume
$3\, {\rm cm}^{-3}$ and $3~{\rm \mu G}$, which are reasonable given
the large uncertainties at the Galactic Center~\cite{Lacroix:2014eea}.
We use the radiation density of ISRF photons included with {\sc
  GALPROP v50}~\cite{galprop,Porter:2008ve}. Since our FOV is $\pm
0.5$ kpc of the Galactic Center, we use the value of the ISRF energy
spectrum tabulated for $R=0,Z=0.25~{\rm kpc}$ in {\sc GALPROP v50} as
the average over the region contributing to the IC flux.

Our final estimates for the IC and bremsstrahlung excesses are,
\begin{eqnarray}
E\frac{dN_{\rm IC,Br}}{dE} = \frac{\rm FOV}{4 \pi} \ell_{\rm IC,Br} \int_{E_{\rm min}}^{m} dE_e
    \left\langle\frac{dn_e}{dE_e}\right\rangle
    \left\langle\frac{dP_{\rm IC,Br}}{dE}\right\rangle\, \nonumber
\end{eqnarray}
where $\langle dP_{\rm IC,Br}/dE\rangle$ are computed using the
average ISRF and gas densities and $\langle dn_e/dE_e\rangle$ is
averaged over the inner 0.5 kpc (in keeping with the small FOV). The
factors $\ell_{\rm IC}$ and $\ell_{\rm Br}$ depend on the details of
the deprojected ISRF and gas densities. For a consistent solution we
expect them to be $O({\rm kpc})$. 
 
 The GCE, IC and bremsstrahlung spectra in the case of a minimal
``democratic'' $e^\pm:\mu^\pm:\tau^\pm=1:1:1$ annihilation channel is
shown in Fig.~\ref{spectraDM} for particle mass $m_{\chi}=8$ GeV and
annihilation cross section $\langle\sigma v\rangle = 3.6 \times
10^{-26}\, {\rm cm}^3{\rm s}^{-1}$. The best-fit dark matter mass when
fitting to the GCE excess is closer to 7 GeV. In this model, the gamma
rays from the $\tau^\pm$ dominate the prompt flux and explain the GCE
excess. In Fig.~\ref{spectraDM}, we have shown IC and bremsstrahlung
model spectra using $\ell_{\rm Br}=\ell_{\rm IC}=1.3\, {\rm kpc}$. We
caution the reader that no attempt has been made to fit to all three
components simultaneously.

The value of the cross section used to create the spectra in
Fig.~\ref{spectraDM} is ruled out by AMS-02 constraints on WIMP
annihilation to leptonic channels~\cite{Bergstrom:2013jra}. However,
the required cross section depends sensitively on the assumed density
profile. For example, if we assume a scale radius ($r_s$) of 10 kpc
and $\gamma=1.2$ (keeping the local density unchanged at $0.3 \, {\rm GeV}/{\rm cm}^3$), the required cross section is a factor of 10
smaller. We have checked that such a profile is consistent with the
expectation that the dark matter halo undergoes adiabatic
contraction~\cite{1986ApJ...301...27B} due to the formation of the
disk and bulge of stars. On the particle physics side, some of this
tension may be relieved by considering annihilation through a vector
mediator, which softens the final $e^\pm$ spectrum.

\section{Conclusions}
We have detected a new excess in gamma
rays toward the Galactic Center that spatially traces starlight
intensity. The spectrum of this new source is consistent with that
produced by high energy electrons and positrons with energies up to
about 10 GeV, upscattering starlight. The population of electrons and positrons required to produce such an inverse Compton emission would
also produce bremsstrahlung radiation due to interactions with the
dense gas at the Galactic Center. Further studies are required to examine the physical implications of this high energy electron population and perform more detailed modeling of the predicted IC excess morphology and spectrum. We detect a gamma-ray excess tracing
20 cm radio map and show that its flux spectrum is consistent in both
shape and amplitude with bremsstrahlung radiation from the same
population of electrons and positrons. We show that the Galactic
Center extended excess that peaks around 2 GeV is also detected at
high significance and that a dark matter model with annihilation to
leptons may provide a consistent explanation for all three excesses. 

While this consistency with three excesses in terms of signal strengths, spectra and morphologies 
is remarkable, other astrophysical explanations exist that deserve detailed investigations. 
Infrared, radio and gamma ray data as incorporated in this work has allowed 
complicated high-energy emissions seen toward the Milky Way's
Galactic Center to be disentangled, and this multiwavelength approach 
may help us to further elucidate the true origin of these excesses.

\vskip 1 cm 

\centerline{\bf Appendix}

\appendix
\section{Robustness of results against background model systematics}
In order to confirm the robustness of our detection, we wish to characterize the behavior of the newly reported IC excess through a range of systematic uncertainties in our analysis. We use the GALPROP code \cite{galprop} to generate additional diffuse background models, and repeat our analysis using these diffuse models in place of the Fermi Pass 7 diffuse model. 

We use models selected from the collection tested by \cite{Calore:2014xka} in their systematic analysis of the GCE signal. For consistency, we will use the same model names as Ref. \cite{Calore:2014xka}. If one examines the IC and $\pi^0$+bremsstrahlung components of the diffuse models in Ref. \cite{Calore:2014xka}, it is apparent that their spectral shapes (with a few exceptions, see model ``E" below) are similar between models and thus running the full suite of 128 backgrounds would produce many degenerate results. Instead, we test two models chosen for their extreme IC parameters, along with two models with more standard parameters. The extreme models were chosen with the intent of testing backgrounds with IC components that were more likely to absorb the excess we have identified and decrease its detection significance. When performing fits using these GALPROP-generated models, we allow the $\pi^0$+bremsstrahlung and IC components to be scaled separately to allow for the possibility that our claimed IC excess is simply an underfitting of the background IC emission. The following models were tested.

\begin{itemize}
\setlength\itemsep{.1cm}
 \item Model A is the reference model in Ref. \cite{Calore:2014xka}.
 \item Model E has a low diffusion coefficient $K_0$. We chose Model E as an extreme test case because the lower value of $K_0$ results in a large increase in the diffuse background IC component at energies below a few GeV and also results in a bump in the combined $\pi^0$+bremsstrahlung component at $\sim$2 GeV. One would expect that these particular spectral features would give `E' the best chance to absorb the new lower-energy IC component as well some of the GCE component at $\sim$2 GeV. 
 \item Model F is the best fit background model from Ref. \cite{Calore:2014xka}.
 \item We also test a model identical to `F' but with optical and infrared ISRF normalization factors raised by 50\%. For this case, we do not allow the $\pi^0$+bremsstrahlung and IC normalizations to float separately.
\end{itemize}

\begin{figure}[h]
\centerline{\includegraphics[width=7.5truein]{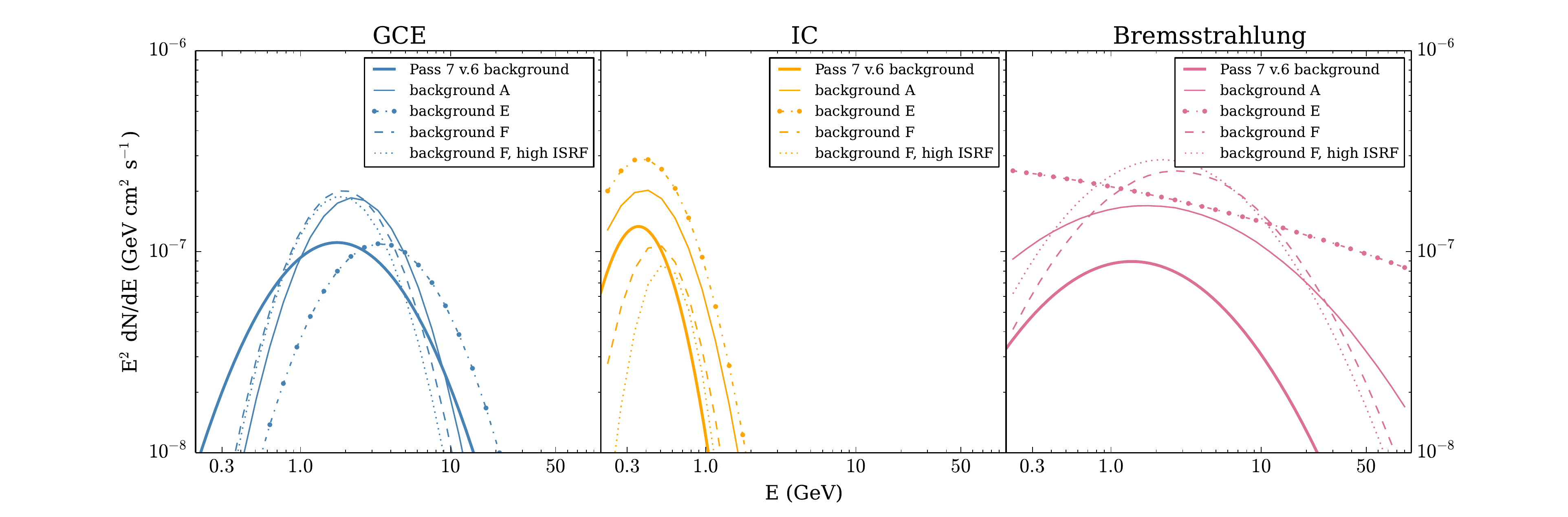}}
\caption{Best-fit parameterized log-parabola model spectra for the GCE, IC, and bremsstrahlung components for tests with varying diffuse background models. $E^2 dN/dE$ is plotted for fluxes \textit{within} the $7^\circ\times7^\circ$ ROI. {\tt ULTRACLEAN} photons were used for all analyses shown. \label{diffusetest}}
\end{figure}

In all our tests, we detect the GCE, IC, and bremsstrahlung templates with similarly high test statistics as when the Fermi Pass 7 background model was used (see Fig. \ref{diffusetest}, Tab. \ref{TSvals}). The spectral shapes of the GCE and IC components are similar for all cases tested. The bremsstrahlung component's spectrum displayed similar features in all background tests with the exception of model E; its spectrum in that case is a power law with no cutoff.

\begin{table*}[h]
\centering
\begin{tabular}{l|ccc}
\hline\hline\\ 

 &\multicolumn{3}{c}{~~Test statistic values for extended sources~~} \\ \\
 ~~Background model & ~~~~~~IC~~~~   & ~~~~~~~GCE~ & ~~~Bremsstrahlung~~ \\ \hline
 \\
 ~~Pass 7             & ~~197  & ~~~~~~207 & 97  \\
 ~~Model A            & ~~386  & ~~~~~~388 & 807  \\
 ~~Model E            & ~~336  & ~~~~~~190 & 1562  \\
 ~~Model F            & ~~123  & ~~~~~~277 & 1707   \\
 ~~Model F, high ISRF ~~~& ~~102  & ~~~~~~242 & 2992   \\ 
 \hline\hline
\end{tabular}
\caption{\label{TSvals} Test statistic values for the IC, GCE, and bremsstrahlung extended sources in each of the diffuse background model tests. The bremsstrahlung template TS values are much higher in the four GALPROP-generated background cases than when the Pass 7 background was used. This may be due to the fact that the Pass 7 model constructs the $\pi^0$+bremsstrahlung component by fitting to gamma-ray data in concentric regions (as opposed to using output maps from GALPROP), which might lead to the absorption of some of the excess bremsstrahlung into the background fit. }
\end{table*}

\begin{figure}[h]
\centerline{\includegraphics[width=5.truein]{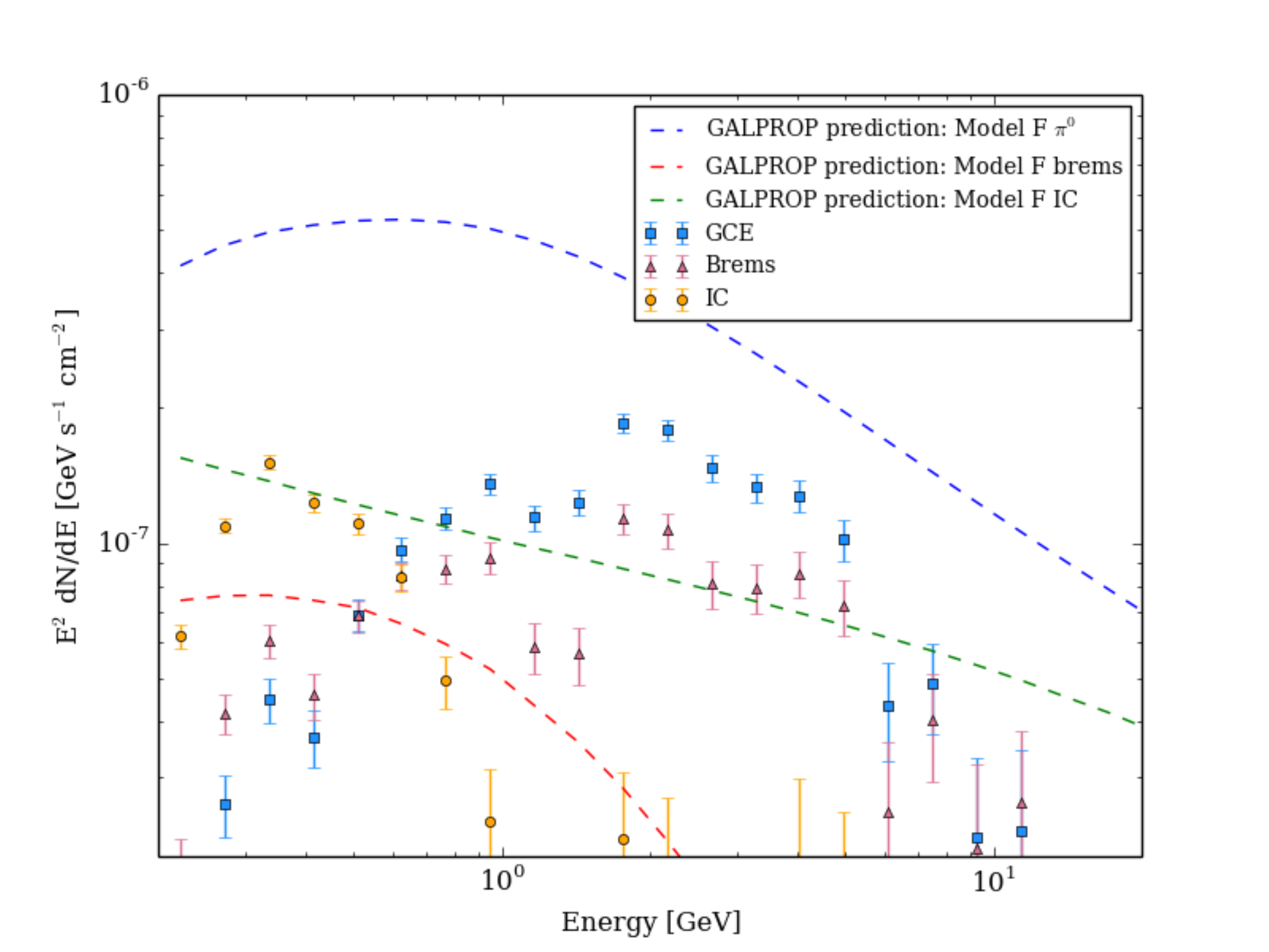}}
\caption{The three diffuse background model components ($\pi^0$/bremsstrahlung/IC) plotted along with the residual spectra for the GCE/IC/bremsstrahlung spectra from the main body of the paper. This plot shows that the excesses are comparable to the relevant backgrounds at their peak energies.\label{modelFcomponents}}
\end{figure}

An additional concern could be that the IC excess flux is very small compared to the flux of the expected IC background contained within the galactic diffuse model. If that were the case, then the IC excess might simply be attributable to uncertainties in the background IC modelling. To check whether this is a relevant concern, we plot in Fig. \ref{modelFcomponents} the IC, GCE, and bremsstrahlung excess component data points from Fig. 2 of the main body of this paper alongside the {\tt GALPROP} predictions for the background $\pi^0$, bremsstrahlung, and IC emission in model F. We show the seperated components of background model F instead of the Pass 7 diffuse model in Fig. \ref{modelFcomponents} because the latter is not generated using {\tt GALPROP} making it difficult to extract the individual $\pi^0$ and bremsstrahlung components for comparison. 

From Fig. \ref{modelFcomponents} it is evident that the IC excess is not a small fraction of the background IC component, but is of the same order of magnitude at its peak. Under similar inspection, the excess bremsstrahlung and GCE normalizations are of the same order as the {\tt GALPROP} predicted bremsstrahlung background, while being $\sim$2--3 times lower than the $\pi^0$ background at their respective peaks. In terms of both spatial and spectral morphology, the three GCE, bremsstrahlung, and IC components differ sufficiently from the diffuse background model such that they can be disentangled through the template likelihood analysis. We conclude that it is unlikely that the IC, GCE and bremsstrahlung excesses might be absorbed within the background model uncertainties.

\begin{figure}[h]
\centering
\includegraphics[width=5.5truein]{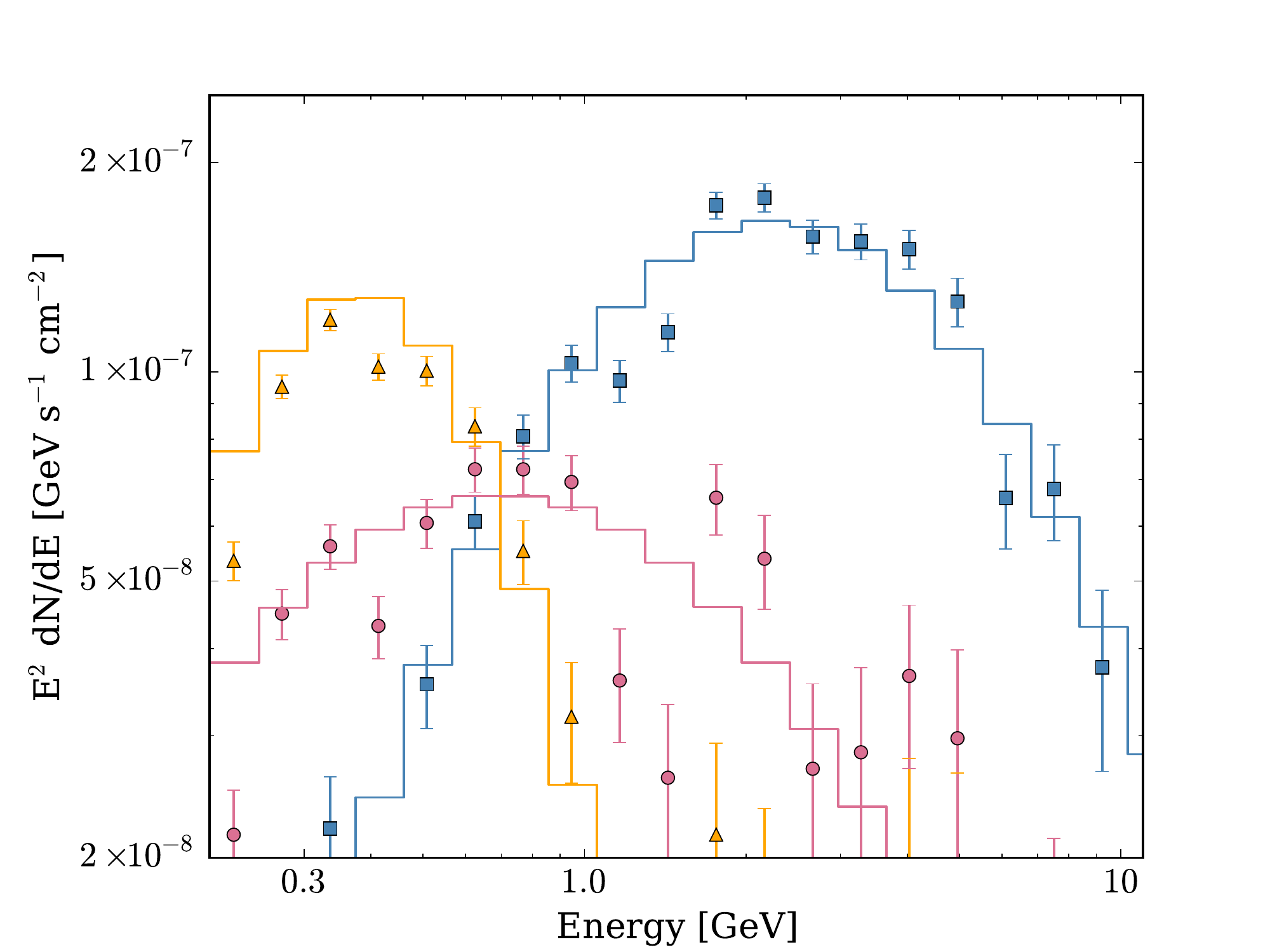}
\caption{Same as Fig. \ref{spectra} but with the HESS residual map substituted in place of the 20 cm radio map for use as the bremsstrahlung template. Note the lower energy cutoff in the HESS template spectrum (pink circle).
\label{hessMGspectra}}
\end{figure}

As the HESS collaboration's Galactic Ridge residuals at TeV energies have also been found to be well-correlated with the morphology at GeV energies \cite{YusefZadeh:2012nh}, we also replaced the 20 cm map with a template based on the HESS residuals \cite{HESScollab:2006}. We recover a spectrum for the HESS template with similar normalization to the 20 cm template spectrum but with a slightly lower energy cutoff. The IC and GCE components' spectra remain consistent with results obtained with the 20 cm template (Fig. \ref{hessMGspectra}). 

The ND component was introduced in Ref. \cite{Abazajian:2014fta} on the GeV galactic center excess as part of a series of checks on how variations in modeling the data might affect the derived properties of the GCE. 
As the ND component was best fit with a spectrum essentially the same as that of the MG template, we interpreted this finding as the ND template tracing molecular gas that had not been captured by the galactic diffuse model and which lay outside the bounds of the MG template. The fact that the best fit ND spatial template is radially \textit{increasing} is also reassuring as it allows for the ND to `fill in' for molecular gas not included in either the MG or Pass 7 model templates, while avoiding overproducing model photons in the very central regions. 

Ref. \cite{Abazajian:2014fta} also found that the Fermi isotropic background template in the $7^\circ\times7^\circ$ had too soft of a spectrum and too low of normalization to adequately describe the isotropic component in the region. To account for this, a `new isotropic' source was included in the analysis (referred to as I in Ref \cite{Abazajian:2014fta}) which was best fit with a hard spectrum dN/dE$\sim E^{-2}$ and detected at high significance. To account for degeneracies between thet new isotropic source I and the Fermi isotropic background, the normalization of the Fermi isotropic source was fixed to unity in Ref. \cite{Abazajian:2014fta}. If the source I was not included in the fit, this excess of isotropic (within the ROI) hard photons was absorbed into the Fermi isotropic background by increasing its overall normalization, with other source components remaining roughly unchanged. This paper does not include the hard isotropic component I. Our best fit for the ND component is spectrally harder (close to dN/dE$\sim E^{-2}$) and has a shallower spatial profile ($\sim\theta^{~0.3}$ as opposed to $\sim\theta^{~0.5}$) than that found in Ref. \cite{Abazajian:2014fta}.  We thus interpret the ND here as a component which can absorb any isotropic or very smoothly distributed photons which are not accounted for in other parts of the model; in this paper these consist of (1) an isotropic component with a harder spectrum and higher amplitude than the standard Fermi template and (2) smoothly distributed gas which lies outside the boundary of the 20 cm template and is not picked up by the galactic diffuse model. If we include a hard isotropic component in lieu of the ND template, our main results for the IC, brems, and GCE components still stand.  

\begin{figure}[h]
 \centering
 \includegraphics[width=5truein]{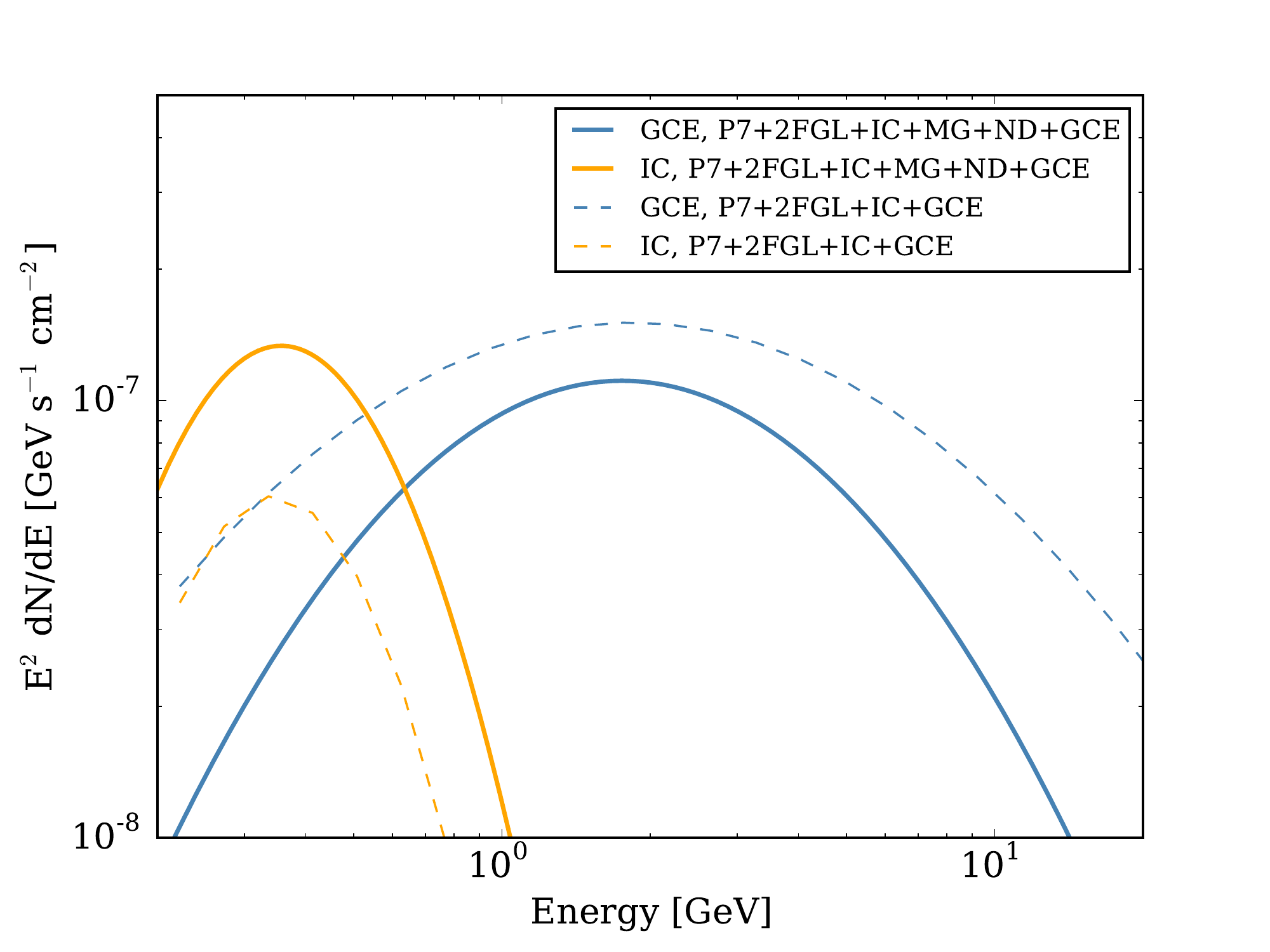}
 \caption{Spectra for the IC (orange) and GCE (blue) components, for cases with full model as described in the paper (P7+2FGL+IC+MG+ND+GCE, solid lines) and with the full model minus the MG and ND components (P7+2FGL+IC+GCE, dashed line). \label{noMGND}}
\end{figure}

Because of our interpretation of the ND component as an extension of the MG template, we perform a fit excluding both ND and MG templates from the full model and verify that our results for the IC and GCE spectra remain consistent with the analysis in the body of the paper. The GCE and IC spectral shapes and normalizations are shown in Fig. \ref{noMGND}; the two cases including and excluding the MG and ND components have have best fit spectra consistent within the background model dependent variance demonstrated in Fig. \ref{diffusetest}.

\clearpage
\acknowledgments
We thank Farhad Yusef-Zadeh for providing the 20 cm
radio maps and Aaron Barth for discussions regarding infrared maps.  
K.N.A. and N.C. are partially supported by NSF
CAREER Grant No. PHY-11-59224, and S.H. by a JSPS fellowship for
research abroad. M. K is partially supported by NSF Grant No. PHY-12-14648.
A.K. is supported by NSF GRFP Grant No. DGE-1321846.

\bibliography{master}

\end{document}